\documentstyle[twoside,fleqn,epsfig]{article}

\def\be{\begin{equation}}
\def\ee{\end{equation}}
\def\bea{\begin{eqnarray}}
\def\eea{\end{eqnarray}}
\newcommand{\lsim}{\mathrel{\mathop{\kern 0pt \rlap
  {\raise.2ex\hbox{$<$}}}
  \lower.9ex\hbox{\kern-.190em $\sim$}}}
\newcommand{\gsim}{\mathrel{\mathop{\kern 0pt \rlap
  {\raise.2ex\hbox{$>$}}}
  \lower.9ex\hbox{\kern-.190em $\sim$}}}

\newcommand{\AmS}{{\protect\the\textfont2
  A\kern-.1667em\lower.5ex\hbox{M}\kern-.125emS}}
\normalsize
\begin{document}
\begin{flushright}
\large
{\bf ROM2F/2008/02 \\}
{\bf submitted for publication\\}
\end{flushright}

\normalsize


\vspace*{0.2cm}

\begin{center}
\Large \bf
Investigation on light dark matter \\
\rm

\vspace{0.8cm}
\large

R.\,Bernabei$^{a,b}$, 
P.\,Belli$^b$, 
F.\,Cappella$^{c,d}$, 
R.\,Cerulli$^{e}$,
C.J.\,Dai$^f$,
H.L.\,He$^f$,
A.\,Incicchitti$^{d}$,
H.H.\,Kuang$^f$,
J.M.\,Ma$^f$,
X.H.\,Ma$^f$,
F.\,Montecchia$^{a,b}$, 
F.\,Nozzoli$^{a,b}$,
D.\,Prosperi$^{c,d}$,
X.D.\,Sheng$^f$, 
Z.P.\,Ye$^{f,g}$,
R.G.\,Wang$^f$, 
Y.J.\,Zhang$^f$

\end{center}

\vspace{1mm}
\noindent $^a${\it Dip. di Fisica, Universit\`a di Roma ``Tor Vergata'', I-00133 Rome, Italy}

\vspace{1mm}
\noindent $^b${\it INFN, sez. Roma ``Tor Vergata'', I-00133 Rome, Italy}

\vspace{1mm}
\noindent $^c${\it Dip. di Fisica, Universit\`a di Roma ``La Sapienza'', I-00185 Rome, Italy}

\vspace{1mm}
\noindent $^d${\it INFN, sez. Roma, I-00185 Rome, Italy}

\vspace{1mm}
\noindent $^e${\it Laboratori Nazionali del Gran Sasso, I.N.F.N., Assergi, Italy}

\vspace{1mm}
\noindent $^f${\it IHEP, Chinese Academy, P.O. Box 918/3, Beijing 100039, China}

\vspace{1mm}
\noindent $^g${\it University of Jing Gangshan, Jiangxi, China}

\normalsize

\begin{abstract}

Some extensions of the Standard Model provide
Dark Matter candidate particles with sub-GeV mass.
These Light Dark Matter particles have been considered for example
in Warm Dark Matter scenarios (e.g.
the keV scale sterile neutrino, axino or gravitino).
Moreover MeV scale DM candidates have been proposed
in supersymmetric models and as source of the 
511 keV line from the Galactic center.
In this paper the possibility of direct detection of a Light Dark Matter candidate
is investigated considering the inelastic scattering
processes on the electron or on the nucleus targets.
Some theoretical arguments
are developed and related phenomenological aspects are discussed.
Allowed volumes and regions for the characteristic phenomenological 
parameters of the considered scenarios are
derived from the DAMA/NaI annual modulation data.

\end{abstract}

{\it Keywords:} Light Dark Matter; underground 
Physics

{\it PACS numbers:} 95.35.+d

\section{Introduction}

Some extensions of the Standard Model provide
Dark Matter (DM) candidate particles with sub-GeV mass;
in the following these candidates will be indicated as 
Light Dark Matter (LDM).

Light Dark Matter particles 
have been considered for example
in Warm Dark Matter scenarios
such as e.g. keV-scale sterile neutrino \cite{sterile}, 
axino or gravitino \cite{axgrav}. 
In addition,
MeV-scale\footnote{It is worth to note that MeV Dark Matter particles can also be considered 
for a possible solution of the missing satellite problem \cite{sss}.}
particles (e.g. axino \cite{MeVax}, gravitino \cite{MeVgrav}, 
heavy neutrinos \cite{rhn,MeVs}, moduli fields 
from string theories \cite{moduli}, {\it Elko} fermions \cite{elko}) 
have been proposed as dark matter
and as source of 511 keV gamma's from the Galactic center, due either to
DM annihilation \cite{dm511a,dm511sv} or to decay\footnote{We also remind that 
the possible decay of DM particles into a lighter state,
with MeV scale splitting, 
has been considered in refs. \cite{dm511d}.}
\cite{MeVax,MeVs} in the bulge.
Moreover, also supersymmetric models exist where  
the LSP naturally has a MeV-scale mass and 
the other phenomenological properties, 
required to generate the 511 keV gammas in the galactic bulge \cite{mevlsp}.

In this paper the direct detection of LDM candidate
particles is investigated considering the possible inelastic scattering
channels either on the electron 
or on the nucleus target. We note, in fact, that -- since the kinetic
energy for LDM particles in the galactic halo does not exceed hundreds eV --
the elastic scattering of such LDM particles both on electrons and on nuclei
yields energy releases well below the energy thresholds of the detectors used in the field;
this prevents the exploitation of the elastic scattering as 
detection approach for these candidates.
Thus, the inelastic process
is the only possible exploitable one for the direct detection of LDM.

\begin{figure} [!ht]
\centering
\includegraphics[width=160pt] {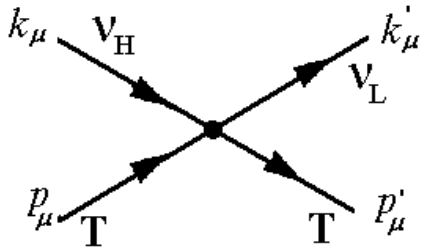}
\caption{Inelastic scattering process considered for the detection of
the Light Dark Matter candidate. The 4-momenta are defined in the laboratory frame. 
The target T can be either an atomic nucleus or an atomic electron.}
\label{fg:diagram}
\end{figure}

The following process is, therefore, considered for detection (see Fig.\ref{fg:diagram}):
the LDM candidate (hereafter named $\nu_H$ with mass $m_H$ and 4-momentum $k_\mu$)
interacts with the ordinary matter target, $T$, with mass $m_T$ and 4-momentum $p_\mu$.
The target $T$ can be either an atomic nucleus or an atomic electron
depending on the nature of the $\nu_H$ particle interaction. 
As result of the interaction a lighter particle is produced
(hereafter $\nu_L$ with mass $m_L < m_H$ and 4-momentum $k'_\mu$) 
and the target recoils with an energy $E_R$, which can be detectable
by suitable detectors.

The lighter particle $\nu_L$ is neutral and it is required 
that it interacts very weakly with ordinary matter;
thus, the $\nu_L$ particle escapes the detector.
In particular, the $\nu_L$ particle can also be another DM halo component 
(dominant or subdominant with respect to the $\nu_H$ one), or 
it can simply be a Standard Model particle (e.g. $\nu_L$ can be identified with
an active neutrino in the scenario of the MeV axino of ref. \cite{MeVax}, where 
the diagrams $\tilde{a}+e^- \rightarrow \nu_{\mu,\tau} + e^-$
have been considered).

The production mechanism of the $\nu_H$ 
(and eventually of the $\nu_L$) in the early Universe
are beyond the scope of the present paper and can be 
investigated in the future; however, since $m_H > m_L$, it is important 
to require that the lifetime of the possible decay of the $\nu_H$ particle
is longer than the age of the Universe. 
In particular, we note that 
the detection process of  Fig.\ref{fg:diagram} also offers
a possible (unavoidable\footnote{All the other decay channels
requires the knowledge/assumptions of some other particle
couplings.}) decay channel 
of $\nu_H$ into $\nu_L$ and $T\bar{T}$ pair
if $m_H > m_L + 2m_T$.
  
Finally, we remind that a LDM particle detection has already been 
discussed in ref. \cite{ijma}, where the possible light bosonic 
(keV mass axion-like particles) Dark Matter candidate has been studied,
considering the total conversion of its mass into electromagnetic radiation.
In the present case, because of the presence of the $\nu_L$ particle in the final state, 
the LDM can be either a boson or a fermion.

In the following, the direct detection of LDM candidate particles is investigated,
some theoretical arguments are developed and
related phenomenological aspects are discussed.
In particular, the impact of these DM candidates will also be discussed in a
phenomenological framework on the basis of the 6.3 $\sigma$ C.L. DAMA/NaI
model independent evidence for particle Dark Matter in the galactic
halo \cite{RNC,ijmd}. We remind that various corollary analyses, considering
some of the many possible astrophysical, nuclear and particle Physics scenarios,
have been analysed by DAMA itself \cite{ijma,RNC,ijmd,epj06,ijma2,chann,wele},
while several others are also available in literature, such as e.g.
refs. \cite{Bo03,Bo04,Botdm,khlopov,Wei01,foot,Saib,droby,sneu,zoom}.
Many other scenarios can be considered as well.
At present the new second generation DAMA/LIBRA set-up is running at the Gran Sasso
National Laboratory of the I.N.F.N..

\section{Detectable energy in the inelastic scattering}

In this section the recoil energy, $E_R$, of the target $T$ 
for the process of Fig.\ref{fg:diagram} is evaluated.

Considering a Lorentz transformation 
with $\vec{\beta} = \frac{\vec{k}+\vec{p}}{k_0+p_0}$
(hereafter $k_0$, $k'_0$, $p_0$ and $p'_0$ are the time components of the respective 4-momenta
in the laboratory frame, see Fig.\ref{fg:diagram}),
it is possible to evaluate the
energy conservation relation in the
center of mass (CM) frame:
\begin{eqnarray}
\sqrt{s}=k_{0,CM}+p_{0,CM} = k'_{0,CM} + p'_{0,CM} \; .
\end{eqnarray}
Knowing that $|\vec{k'}_{CM}| = |\vec{p'}_{CM}|$, one obtains: 
\begin{eqnarray}
p'_{0,CM} = \frac{m_T^2-m_L^2}{2 \sqrt{s} } + \frac{ \sqrt{s} }{2} =
\sqrt{p'^2_{CM} + m_T^2} \; .
\end{eqnarray}
Defining the
Lorentz boost factor:
$\gamma = 1/\sqrt{1-\beta^2} = \frac{k_0+p_0}{\sqrt{s}}$
one can write the total energy of the target in the laboratory frame
after the scattering by means of a Lorentz transformation:
\begin{eqnarray}
p'_0 = \gamma \left( p'_{0,CM} + \vec{\beta} \cdot \vec{p'}_{CM} \right)
\end{eqnarray}
Assuming the target at rest before the scattering, i.e. $\vec{p}=0$ and $p_0=m_T$, and
the non-relativistic nature of the LDM particle, i.e. $k_0 \simeq m_H$ and $k \simeq m_H v_{LDM}$,
with $v_{LDM} \sim 10^{-3}c$,  
the energy released to the target $T$:
\mbox{$E_R = k_0-k'_0 = p'_0 - p_0$} 
(that is, the target recoil energy) can be written as:
\begin{eqnarray}
E_R &=& p'_0 - m_T 
= (\gamma p'_{0,CM} - m_T) + \gamma \beta p'_{CM} cos \theta_{CM} \\
&=& \langle E_R \rangle + \frac{E_+-E_-}{2} cos \theta_{CM}
\label{eq:ED}
\end{eqnarray}
where:
i)   $\theta_{CM}$ is the scattering angle in the CM frame;
ii)  $\langle E_R \rangle$ is the average energy; 
iii) $E_-$ is the minimum recoil energy;
iv)  $E_+$ is the maximum recoil energy. 

Considering the non-relativistic nature of the LDM, one gets:
$s \simeq \left( m_T +  m_H \right)^2$, and for the $ \langle E_R \rangle$ average energy:
\begin{eqnarray}
\langle E_R \rangle = \frac{E_++E_-}{2} \simeq \frac{\bar{m} \; \Delta}{m_H+m_T} \;.
\label{eq:ED3}
\end{eqnarray}
There $\Delta = m_H-m_L$ is the mass splitting of the $\nu_H$-$\nu_L$ particle system and
$\bar{m} = \frac{m_H+m_L}{2}$ is the average mass of the $\nu_H$-$\nu_L$ particle system.

The spread of the recoil energy:
\begin{eqnarray}
\frac{E_+-E_-}{ \langle E_R \rangle } \simeq k \sqrt
{
\frac{8m_T}{\bar{m} \; \Delta (m_H + m_T)}
}
\label{eq:ED4}
\end{eqnarray}
can be appreciable only for hundreds MeV LDM interacting on heavy nucleus targets. 
As an example, assuming $m_H=1$ MeV and $\Delta=0.5$ MeV, the spread in energy
is few $10^{-3}$ both for electrons and for every nuclear targets. 
Therefore, since the LDM kinetic energy is expected to be negligible with 
respect to $\Delta$, one has:
$ \langle E_R \rangle \gg (E_+-E_-)$, and the recoil energy is 
approximatively fixed at the $ \langle E_R \rangle $ value.

\begin{figure} [!t]
\centering
\includegraphics[width=220pt] {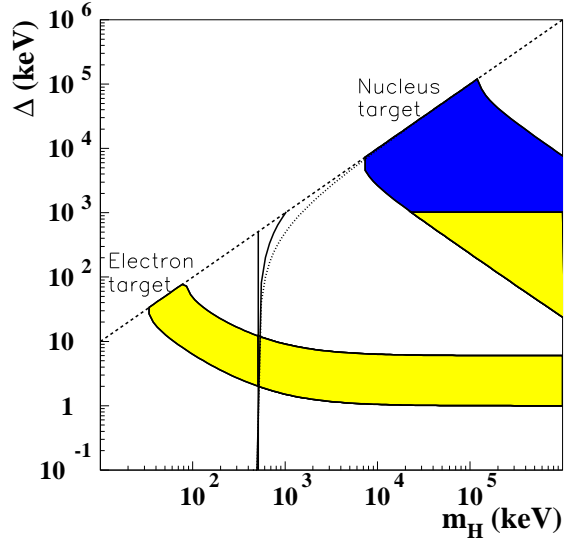}
\caption{Configurations in the plane $\Delta$ vs $m_H$ (shaded areas), corresponding to 
released energies in NaI(Tl) within the energy interval 1-6 keV electron equivalent.
The upper area is due to LDM interaction on Na and I nuclear targets, while the lower 
area to LDM interaction on electron target.
The dashed line ($m_H = \Delta$) marks the case where $\nu_L$ is a massless particle.
The configurations characterized by $\Delta \ge 2m_e$ (dark area)
are of interest for the positron annihilation line
from the galactic center through the decay: $\nu_H \rightarrow \nu_L e^+e^-$.
The thresholds of the possible annihilation processes: 
$\nu_H \bar{\nu}_H \rightarrow e^+ e^- $ (solid vertical line at $m_H$ = 511 keV);
$\nu_H \bar{\nu}_L \rightarrow e^+ e^- $ (solid curve);
$\nu_L \bar{\nu}_H \rightarrow e^+ e^- $ (solid curve); 
$\nu_L \bar{\nu}_L \rightarrow e^+ e^- $ (dotted curve) are shown.
}
\label{fg:reg}
\end{figure}

\vspace{0.3cm}

In order to offer to the reader just a view of the detectability of LDM particles,
Fig. \ref{fg:reg} shows examples of the configurations in the plane $\Delta$ vs $m_H$ (shaded areas), 
corresponding to released energies in NaI(Tl) within the energy interval 1-6 keV electron equivalent.
The lower area in Fig. \ref{fg:reg} refers to LDM interacting on electron targets, while
the upper area refers to LDM interacting on the Na and I targets 
(the effect of the quenching factor \cite{RNC} and of the channeling in the NaI lattice 
\cite{chann} have been considered).
The dashed line $m_H = \Delta$ marks the configurations where $\nu_L$ is a massless particle
(or also a very light particle, such as e.g. an active neutrino or a nearly massless sterile 
one or the light axion, etc.).
The configurations characterized by $\Delta \ge 2m_e$ (dark area; hereafter $m_e$ is the electron mass) 
are of interest for the positron annihilation line
from the galactic center due to possible (energetically
allowed) decay: $\nu_H \rightarrow \nu_L e^+e^-$.
Finally, the configurations on the right of the vertical line at $m_H$ = 511 keV
are of interest for the possible annihilation process:
$\nu_H \bar{\nu}_H \rightarrow e^+ e^- $ in the galactic center, while
the configurations of $m_H$ on the right of the solid or dotted curves are of  
interest for the possible annihilation processes:
$\nu_H \bar{\nu}_L \rightarrow e^+ e^- $ (solid curve),
$\nu_L \bar{\nu}_H \rightarrow e^+ e^- $ (solid curve),
and 
$\nu_L \bar{\nu}_L \rightarrow e^+ e^- $ (dotted curve) in the galactic center.

\section{Interaction rate of LDM particle}

The interaction rate of the LDM particle with the target $T$ and for the studied process
can be written as:
\begin{equation}
\frac{dR_T}{dE_R} = 
\eta_T \frac{\rho_{\nu_H}}{m_H} 
\int \frac{d\sigma_T}{dE_R}
v f(\vec{v})d^3v \; ,
\label{eq:rate}
\end{equation}
where: 
i)   $\rho_{\nu_H} = \xi \rho_0$, with $\rho_0$
     local halo density and $\xi \leq 1$ fractional amount 
     of $\nu_H$ density in the halo; 
ii)  $\eta_T$ is the target number density in the detector; 
iii) $\frac{d\sigma_T}{dE_R}$ is the differential cross section of the considered LDM particle inelastic scattering
     with the target $T$;
iv)  $f(\vec{v})$ is the velocity ($\vec{v}$) distribution of the $\nu_H$ particles 
     in the Earth (laboratory) frame.  
 
Since the sub-GeV LDM wavelength 
($\lambda = \frac{h}{k} > 10^3$ fm) is much larger than the nucleus 
size, the targets can be considered as point-like and the form factors of 
the targets can be
approximated by unity. In such a reasonable hypothesis and assuming, for simplicity,
the isotropy of the differential cross section\footnote{
Different assumptions would produce quite similar phenomenologies, since -- as mentioned --
\mbox{$\langle E_R \rangle \gg (E_+-E_-)$}, and the recoil energy is 
approximatively fixed at the $\langle E_R \rangle$ value.}, one gets:
\begin{equation}
\frac{d\sigma_T}{dE_R} \simeq \frac{\sigma_T}{E_+-E_-} 
\Theta(E_+-E_R)
\Theta(E_R-E_-) \; ,
\label{eq:sv1}
\end{equation} 
where: i) $\sigma_T$ is the target cross section; 
ii) the Heaviside functions, $\Theta$, 
define the domain of the differential cross section.
We note that $E_+$ and $E_-$ depend on the target mass
and slightly on the LDM velocity, $v$.

Generally, $\sigma_T$ can be a function of $v$, depending on
the peculiarity of the particle interaction. In the following,
we adopt a widely considered approximation for the non-relativistic case
\cite{dm511sv,mevlsp,sigmav} (the subscript $T$ drops):
\begin{equation}
\sigma v \simeq a + b v^2 \; .
\label{eq:sigmav}
\end{equation}
There $a$ and $b$ are constant and depend on the peculiarity of the particle interaction
with the target $T$; 
they will be considered as free parameters in the description 
of the process.
In the following, in order to deal the direct detection process with more usual 
parameters, the cross sections $\sigma_0^T = \frac{a}{v_0}$ and 
$\sigma_m^T = b v_0$ will be used as free 
parameters; they are 
related to the $a$ and $b$ parameters rescaled with the Dark Matter local 
velocity, $v_0$, \cite{Ext,loc}.

\noindent Thus, eq. (\ref{eq:rate}), becomes:
\begin{equation}
\frac{dR_T}{dE_R} = 
\eta_T \frac{\rho_{\nu_H}}{v_0m_H} 
\int \left(
\sigma_0^T v_0^2 + \sigma_m^T v^2 \right)
\frac{\Theta(E_+-E_R)
\Theta(E_R-E_-)}{E_+-E_-}  
f(\vec{v})d^3v \; .
\label{eq:rate2}
\end{equation}
Since: $\int_0^\infty \frac{\Theta(E_+-E_R) \Theta(E_R-E_-)}{E_+-E_-} dE_R = 1$,
the total rate $R_T^{tot}$ can be written as:
\begin{equation}
R_T^{tot} = \int_0^\infty \frac{dR_T}{dE_R} dE_R = 
\eta_T \frac{\rho_{\nu_H}}{v_0m_H} 
\left( \sigma_0^T v_0^2 + \sigma_m^T \langle v^2 \rangle \right) \; .
\label{eq:rate2a}
\end{equation}

Similarly as the light bosonic case discussed in ref. \cite{ijma}, 
the $\sigma_m^T \langle v^2 \rangle$ term 
provides an annual modulation of the expected counting rate
for LDM interactions.
In fact, the velocity of the LDM particle in the galactic frame 
can be defined as: 
$\vec{v}_g = \vec{v}+\vec{v}_{\oplus}$. 
Thus, one obtains for non rotating halo:
$\langle v^2 \rangle = \langle v_g^2 \rangle + v_{\oplus}^2 $.
Since the Earth velocity in the galactic frame ($\vec{v}_{\oplus}$) is given by the sum of the
Sun velocity ($\vec{v}_{\odot}$) and of the Earth's orbital time-dependent velocity
around the Sun ($ \vec{v}_{SE}(t)$), neglecting the $v^2_{SE}$ term one gets:
\begin{equation}
\langle
v^2
\rangle \simeq
\langle
v_g^2
\rangle+
v_{\odot}^2
+ 2 \vec{v}_{\odot} \cdot \vec{v}_{SE}(t) \simeq
\langle
v_g^2
\rangle+
v_{\odot}^2
+ v_{\odot} v_{SE} cos (\omega(t-t_0)) \; .
\label{eq:vel32}
\end{equation}
In the last equation the angle of the terrestrial orbit with respect to the galactic plane
has been considered to be $\simeq 60^o$. Moreover, we consider
$\omega=2 \pi/T$ with $T=1$ year, and the phase $t_0$ (which corresponds 
to $\simeq$ 2nd June, that is, when the Earth's speed in the galactic frame is
at the maximum).
The Sun velocity can be written as $ |\vec{v}_{\odot}| \simeq v_0 + 12 $ km/s,
where $v_0$ is the local velocity, whose value is in the range 170-270 km/s
\cite{Ext,loc}.
The Earth's orbital velocity is $ v_{SE} \simeq 30 $ km/s.
Finally, $\langle v_g^2 \rangle$ depends on the halo model and on
the $v_0$ value (just for reference, in the particular simplified case of
isothermal halo model:  $\langle v_g^2 \rangle = \frac{3}{2} v_0^2$).

In conclusion, the expected signal is given by the sum of two contributions:
one independent on the time and the other one dependent on the time through 
\mbox{$cos (\omega(t-t_0))$},
with relative amplitude depending on the adopted scenario.

\vspace{0.2cm}
\subsection{Interaction with atomic electrons}

For atomic electron targets, the energy interval $(E_-, E_+)$ 
is always very narrow $\left( \frac{E_+ - E_-}{\langle E_R \rangle} \lsim few \;\% \right)$. Therefore, 
using the approximation
$\frac{\Theta(E_+-E_R) \Theta(E_R-E_-)}{E_+-E_-} \simeq \delta(E_R-\langle E_R \rangle)$,
the interaction rate with atomic electrons can be simply written as:

\begin{equation}
\frac{dR_e}{dE_R} \simeq
\eta_e \frac{\rho_{\nu_H}}{v_0 m_H} 
\left(
\sigma^e_0  v_0^2+ 
\sigma^e_m
\langle v^2 \rangle
\right)
\delta(E_R-\langle E_R \rangle)
\label{eq:rate3}
\end{equation}
where $\eta_e$ 
is the electron number density
in the target detector 
and $\sigma^e_0$ and $\sigma^e_m$ are the average (over all the atomic 
electrons)
cross section parameters 
for electrons in the target material.

After the interaction the final state can have -- beyond the $\nu_L$ particle -- 
either a prompt electron and an ionized atom or an excited atom plus 
possible X-rays/Auger electrons. 
Therefore, the process produces X-rays and electrons of relatively low energy, 
which are mostly contained with efficiency $\simeq 1$ in a detector of 
a suitable size.

Therefore, the differential counting rate can simply  be written accounting for the
detector energy resolution, by means of the $G(E,E_R)$ kernel, which 
generally has a Gaussian behaviour:
\begin{equation}
\frac{dR_e}{dE} = \int G(E,E_R)\frac{dR_e}{dE_R}dE_R = R_e^{tot} \times \frac{1}{\sqrt{2\pi}\delta}
e^{-\frac{\left(E-\langle E_R \rangle \right)^2}{2\delta^2}} \, .
\label{eq:rate333}
\end{equation}
There $\delta$ is the energy resolution of the detector and $R_e^{tot}$ is the area of the Gaussian
peak centered at the $\langle E_R \rangle$ value.

Finally, the expected differential rate is given by the sum of two contributions:
$\frac{dR_e}{dE} = S_0 + S_m \cdot cos \omega (t-t_0)$, 
where $S_0$ and $S_m$ are the unmodulated and the modulated part of the expected differential 
counting rate, respectively.

\vspace{0.2cm}
\subsection{Interaction with nuclei}

As regards the interaction with target nuclei, when considering 
the cases of heavy target and for relatively ``high'' LDM mass 
the energy interval $(E_-, E_+)$ cannot be neglected.
Therefore, the interaction rate with the nucleus $T$ can be approximated as:
\begin{equation}
\frac{dR_T}{dE_R} \simeq 
\eta_T \frac{\rho_{\nu_H}}{v_0 m_H} 
\left(
\sigma^T_0  v_0^2+ \sigma^T_m
\langle v^2 \rangle
\right)
\frac{
\Theta(E^T_+-E_R)
\Theta(E_R-E^T_-)
}{E^T_+-E^T_-} \; ,
\label{eq:rate4}
\end{equation}
where $\eta_T$ is the nucleus number density in the material 
and $E_{\pm}^T$ are the $E_{\pm}$ values for the nucleus $T$ calculated at $v=\langle v \rangle$.

Considering the case of a crystal detector target (as e.g. the NaI(Tl)), the 
possible channeling effect \cite{chann} has also to be taken into account. 
The detection of a nuclear recoil of kinetic 
energy, $E_R$, is, in particular, related to the response function 
$\frac{dN_T}{dE_{det}}(E_{det}|E_R)$ \cite{chann},
where $E_{det}$ is the released energy in keV electron equivalent.
Thus, the differential $E_{det}$ distribution
for the nucleus $T$ when 
accounting for the channeling effect can be written as: 
\begin{equation}
\frac{dR^{(ch)}_T}{dE_{det}}(E_{det}) = \int
\frac{dN_T}{dE_{det}} (E_{det}|E_{R})
\frac{dR_T}{dE_{R}}(E_{R}) dE_R  \; .
\label{eq:rt}
\end{equation}

Hence, the expected differential counting rate for LDM interaction with nucleus $T$
can easily be derived accounting for the detector energy resolution kernel $G(E,E_{det})$:
\begin{equation}
\frac{dR_{T}^{(ch)}}{dE} =
\int G(E,E_{det}) 
\frac{dR^{(ch)}_T}{dE_{det}} dE_{det} \; .
\label{eq:labelmul}
\end{equation}

Finally, as for the electron interacting LDM, 
the expected differential rate is given by the sum of two contributions:
$\frac{dR_T}{dE} = S_0 + S_m \cdot cos \omega (t-t_0)$, 
where $S_0$ and $S_m$ are the unmodulated and the modulated part of the expected differential 
counting rate, respectively.

\subsection{Some examples on interaction rate in NaI(Tl)}

In this section as template purpose, some examples on interaction rate of LDM
particles in NaI(Tl) detectors are given.
In particular, since the expected energy distribution for 
electron\footnote{Here, we only remind that $\eta_e \simeq 2.6 \times 10^{26}$ kg$^{-1}$ in NaI(Tl).}
interacting LDM has a simple Gaussian behaviour (see eq. (\ref{eq:rate333})),
only some examples for the nucleus interacting LDM will be shown in the following.

In NaI(Tl) detectors, both Na and I nuclei 
($\eta_{Na}=\eta_I \simeq 4 \times 10^{24}$ kg$^{-1}$)
can contribute to the detection of LDM. Therefore, the total expected counting rate is
given by the sum of two contributions similar as eq. (\ref{eq:labelmul}), arising from 
the two nuclei species (T = Na, I). 
The cross section parameters $\sigma^T_0$ and $\sigma^T_m$ for Na and I are related
with some model dependent scaling laws; 
in particular, in the following, the two 
illustrative cases of coherent
\mbox{($\sigma^{coh} \simeq \frac{\sigma^{Na}}{A^2_{Na}} \simeq \frac{\sigma^{I}}{A^2_{I}}$)} 
and incoherent
($\sigma^{inc} \simeq \sigma^{Na} \simeq \sigma^{I}$) nuclear scaling laws are investigated.  
Obviously, other different nuclear scaling laws are in principle possible. 

\begin{figure} [!ht]
\centering
\includegraphics[width=6.cm] {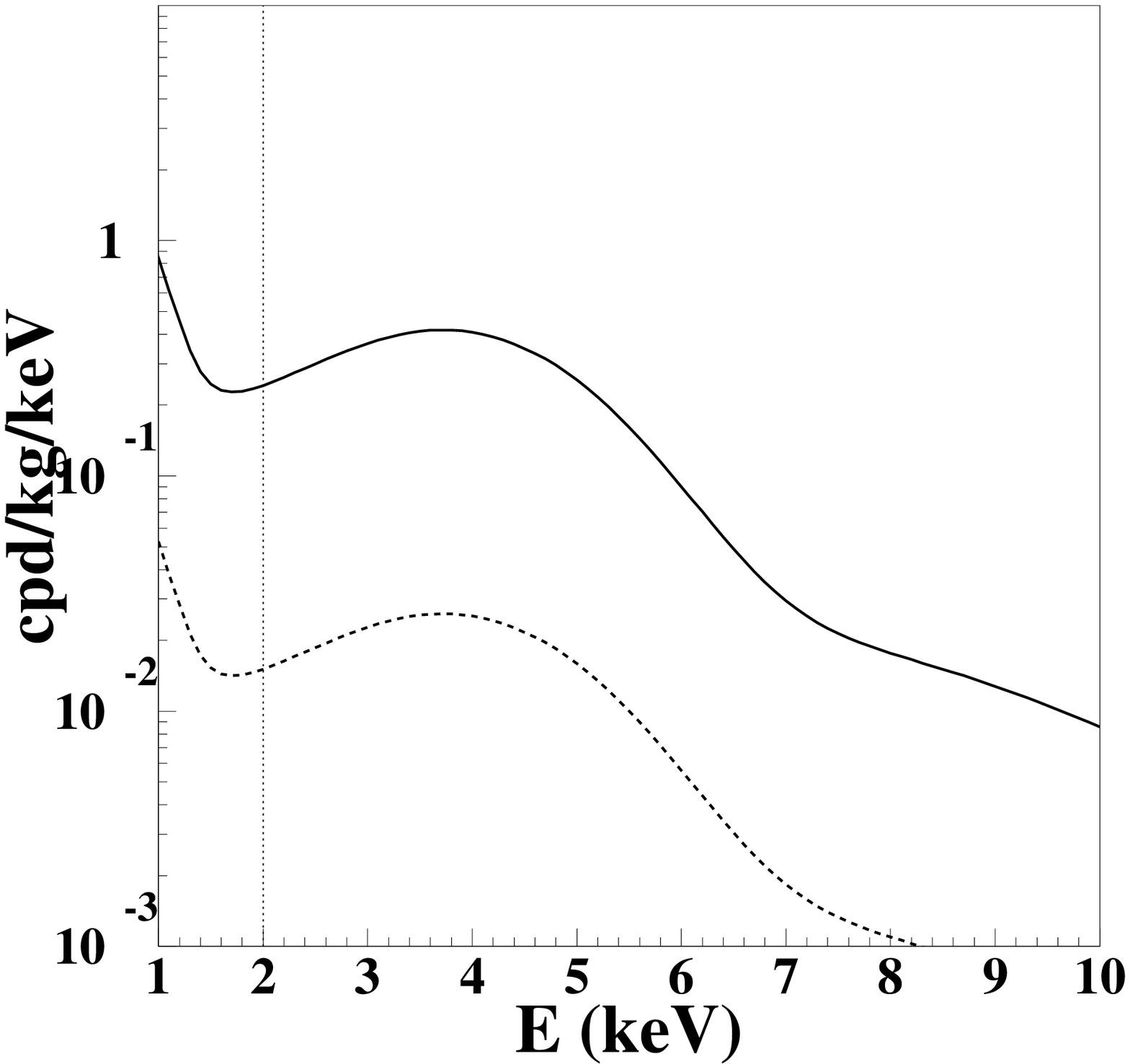}
\includegraphics[width=6.cm] {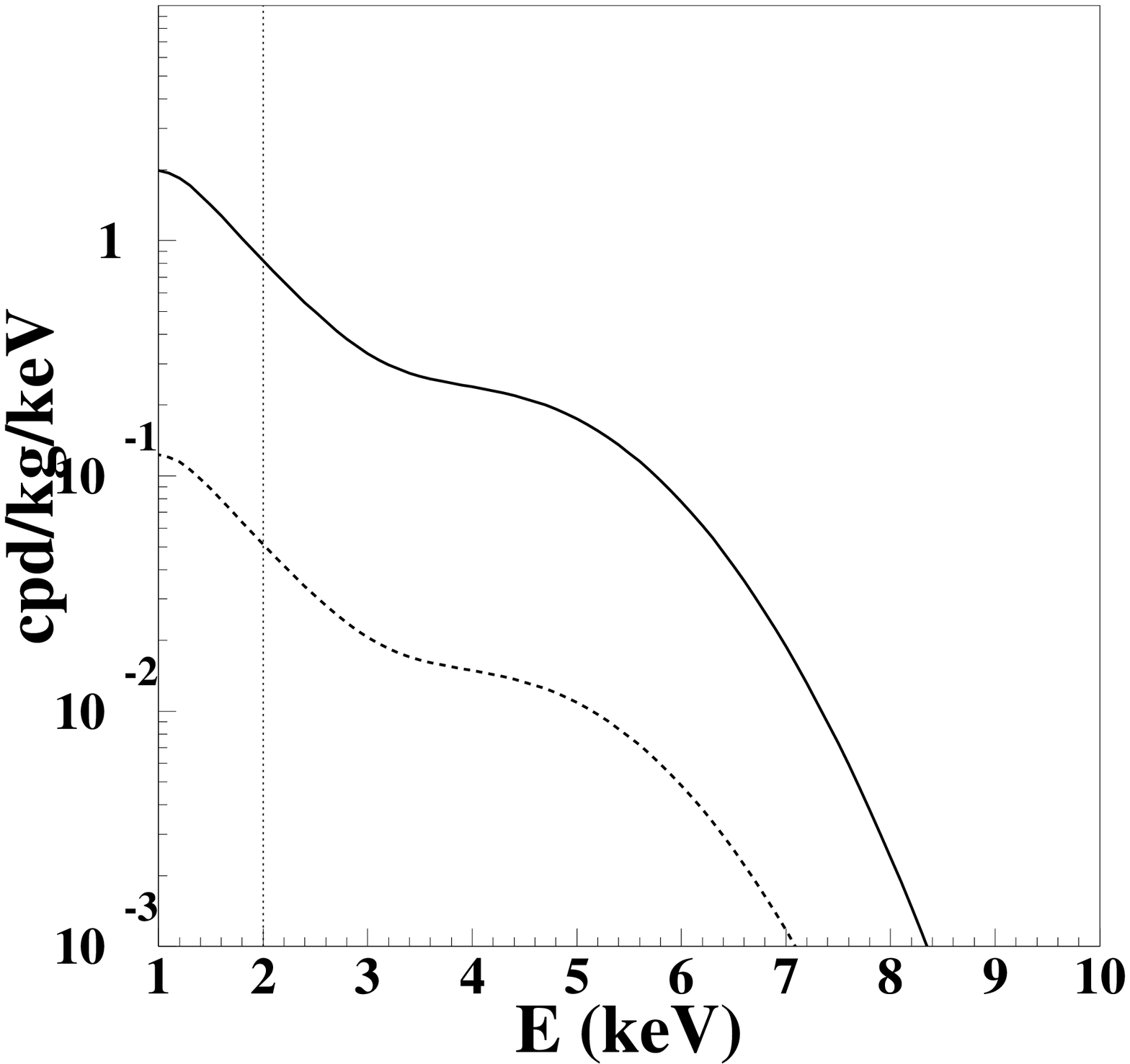}
\caption{Examples of the energy distributions of the unmodulated (solid) and of the 
modulated (dotted) parts of the expected differential rate in NaI(Tl) for interactions
of LDM with $m_H=100$ MeV on the target nuclei. 
{\em Left:} case of coherent cross section scaling laws; here $\Delta = 4.8$ MeV 
and $\xi \sigma_0^{coh} \ll \xi \sigma_m^{coh} \simeq 2 \times 10^{-6}$ pb.
{\em Right:} case of incoherent cross section scaling laws; here $\Delta = 0.95$ MeV 
and $\xi \sigma_0^{inc} \ll \xi \sigma_m^{inc} \simeq 20 \times 10^{-3}$ pb.
The A5 halo model (a NFW halo model with local velocity equal to 220 km/s
and density equal to the maximum value, see ref. \cite{RNC,ijmd}) has been considered. 
The quenching factors have been assumed as the case A of ref. \cite{RNC,ijmd}. 
The channeling effect has been included; see text.
The vertical dotted lines correspond to the energy threshold of the NaI(Tl) detectors
used in DAMA/NaI set-up.}
\label{fg:spectra_dm}
\end{figure}

As template purpose, in Fig. \ref{fg:spectra_dm}, the expected energy distributions of the 
unmodulated/modulated part of the differential rate in NaI(Tl)
are shown for interactions of LDM with $m_H=100$ MeV on the target nuclei. 
In particular, there are shown 
the cases of coherent cross section scaling laws (Fig. \ref{fg:spectra_dm} {\it --left})
with $\Delta = 4.8$ MeV and $\xi \sigma_0^{coh} \ll \xi \sigma_m^{coh} \simeq 2 \times 10^{-6}$ pb
and the case of incoherent cross section scaling laws (Fig. \ref{fg:spectra_dm} {\it --right})
with $\Delta = 0.95$ MeV and $\xi \sigma_0^{inc} \ll \xi \sigma_m^{inc} \simeq 20 \times 10^{-3}$ pb.
The A5 halo model (a NFW halo model with local velocity equal to 220 km/s
and density equal to the maximum value, see ref. \cite{RNC,ijmd}) has been considered. 
The quenching factors have been assumed as the case A of ref. \cite{RNC,ijmd} and 
the channeling effect has been included, as mentioned.

It is worthwhile to note that similar behaviors can also be obtained by using 
other choices of the halo model, quenching factors and values of the masses
of the involved particles in the interaction.

\section{Data analysis and results for LDM candidates in DAMA/NaI}

The 6.3 $\sigma$ C.L. model independent
evidence for Dark Matter particles in the galactic halo
achieved over seven annual cycles by DAMA/NaI \cite{RNC,ijmd}
(total exposure $\simeq 1.1 \times 10^{5}$ kg $\times$ days)
can also be investigated for the case of a LDM
candidate (in addition to the other corollary quests
analyzed by DAMA itself \cite{ijma, RNC,ijmd,epj06,ijma2,chann,wele} and
available in literature, e.g.
refs. \cite{Bo03,Bo04,Botdm,khlopov,Wei01,foot,Saib,droby,sneu,zoom}).

In the following, the same dark halo models
and related parameters given in table VI of ref. \cite{RNC} are considered.
The related DM density is given in table VII of the same reference. 
Note that, although a large number of self-consistent galactic halo models 
have been considered, still many other possibilities exist.
As regards the case of LDM interacting on nuclei, also the Na and I quenching factors \cite{RNC} 
and the Na and I channeling response functions given in \cite{chann} have been considered.
The uncertainties on the Na and I quenching factors have been taken into account as done in ref. \cite{RNC}. 
In addition, the presence of the existing Migdal effect and of the possible SagDEG 
contributions -- we discussed in \cite{ijma2,epj06}, respectively --
will be not included here for simplicity.

As regards the cross section parameters, the $\sigma_0$ terms 
do not contribute to the annual modulation of the signal.
In particular, the measured modulation amplitude easily allows to derive
$\sigma_0 \lsim \sigma_m$.
In the following, for simplicity, the contributions of the $\sigma_m$ terms 
are assumed to be dominant with respect to the 
contributions of the $\sigma_0$ ones. 

The results are calculated by taking into
account the time and energy behaviours of the {\it single-hit}\footnote{The
{\it single-hit} events are those events where only one detector of many actually
fires in a multi-detectors set-up. In particular, the $\nu_H$ particle is no more present
after the inelastic interaction and, therefore, can only provide {\it single-hit} events 
in a multi-detectors set-up.}
experimental data
through the standard maximum likelihood method\footnote{Shortly, the likelihood function is:
${\it\bf L}  = {\bf \Pi}_{ijk} e^{-\mu_{ijk}}
{\mu_{ijk}^{N_{ijk}} \over N_{ijk}!}$, where
$N_{ijk}$ is the number of events collected in the
$i$-th time interval, by the $j$-th detector and in the
$k$-th energy bin. 
$N_{ijk}$ follows a Poissonian
distribution with expectation value
$\mu_{ijk} = [b_{jk} + S_{0,k} + S_{m,k} \cdot  cos\omega(t_i-t_0)] M_j \Delta
t_i \Delta E \epsilon_{jk}$.
The unmodulated and modulated parts of the signal,
$S_{0,k}$ and $S_{m,k}cos\omega(t_i-t_0)$, respectively,
are here functions of the LDM mass $m_H$, the splitting $\Delta$
and the cross section $\sigma_m$.
The b$_{jk}$ is the background contribution;
$\Delta t_i$ is the detector running time during the $i$-th time interval;
$\Delta E$ is the energy bin;
$\epsilon_{jk}$ is the overall efficiency and $M_j$ is the detector mass.}.
In particular, they are presented in terms of slices of the three-dimensional allowed volume
($m_H$, $\Delta$, $\xi\sigma_m$) -- where $\sigma_m$ is alternatively
$\sigma_m^e$, $\sigma_m^{coh}$ and $\sigma_m^{inc}$ -- 
obtained as superposition of the configurations corresponding
to likelihood function
values {\it distant} more than $4\sigma$ from
the null hypothesis (absence of modulation) in each one of the several
(but still a very limited number) of the considered model frameworks.
In this way one accounts for at least some of the
existing theoretical and experimental uncertainties.
It is worth to note that 
the inclusion of other existing uncertainties would further extend the 
allowed volumes/regions and increase the sets of obtained best fit values.

The projection of the whole 4$\sigma$ allowed
volume on the plane ($m_H$, $\Delta$) gives typical patterns
similar as those reported in Fig.\ref{fg:reg}.
Moreover, since the two regions are disconnected,
the LDM detection is always dominated by only one 
of the different target contributions.
Therefore, in the following, slices of the allowed volume either 
for electron interacting LDM or nucleus interacting LDM
are presented separately.

\subsection{Case of electron interacting LDM}

\begin{figure} [!ht]
\centering
\includegraphics[width=7.cm] {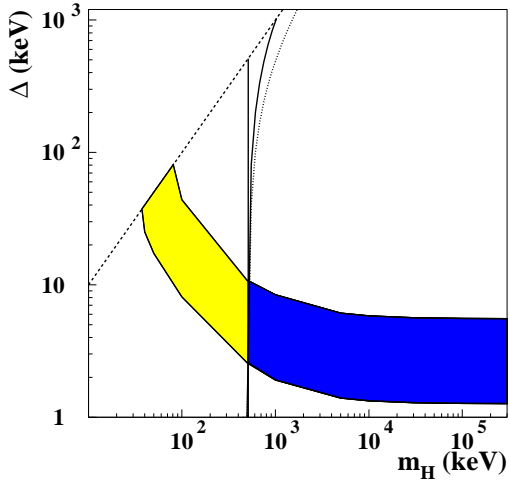}
\caption{Case of electron interacting LDM.
Projection of the 4$\sigma$ allowed 3-dimensional volume on the plane ($m_H$, $\Delta$)
for the same dark halo models and parameters described in ref. \cite{RNC};see text.
The dashed line ($m_H = \Delta$) marks the case where $\nu_L$ is a massless particle.
The decay through the detection channel,
$\nu_H \rightarrow \nu_L e^+ e^-$, is energetically not allowed
for the selected configurations.
The configurations with $m_H \gsim m_e$ (dark area)
are interesting for the possible annihilation processes: 
\mbox{$\nu_H \bar{\nu}_H \rightarrow e^+ e^- $},
$\nu_H \bar{\nu}_L \rightarrow e^+ e^- $,
$\nu_L \bar{\nu}_H \rightarrow e^+ e^- $,
and $\nu_L \bar{\nu}_L \rightarrow e^+ e^- $ 
in the galactic center.
The three nearly vertical curves are the thresholds of these latter processes as
mentioned in the caption of Fig.\ref{fg:reg} and in the text.}
\label{fg:elett1}
\end{figure}

In case of electron interacting LDM, 
the projection of the 4$\sigma$ allowed volume on the plane ($m_H$, $\Delta$)
for the same dark halo models and parameters described in ref. \cite{RNC}
is reported in Fig.\ref{fg:elett1}. 
The allowed $m_H$ values and the splitting $\Delta$ 
are in the intervals \mbox{35 keV $\lsim m_H \lsim $ O(GeV)}\footnote{\label{fn:1} For values of $m_H$ greater than
O(GeV), the definition of Light Dark Matter is no more 
appropriate. Moreover, the kinetic energy of the particle 
would be enough for the detection also through 
the elastic scattering process, as demonstrated in ref. \cite{wele}.}
and \mbox{1 keV $\lsim \Delta \lsim 80$ keV}, respectively.
It is worth to note that in such a case the decay through the detection channel:
$\nu_H \rightarrow \nu_L e^+ e^-$, is energetically forbidden
for the given $\Delta$ range.
The configurations with $m_H \gsim 511$ keV 
(dark area in Fig.\ref{fg:elett1})
are instead of interest for the possible annihilation processes: 
\mbox{$\nu_H \bar{\nu}_H \rightarrow e^+ e^- $},
$\nu_H \bar{\nu}_L \rightarrow e^+ e^- $,
$\nu_L \bar{\nu}_H \rightarrow e^+ e^- $,
and $\nu_L \bar{\nu}_L \rightarrow e^+ e^- $, 
in the galactic center.

\begin{figure} [!ht]
\centering
\includegraphics[width=210pt] {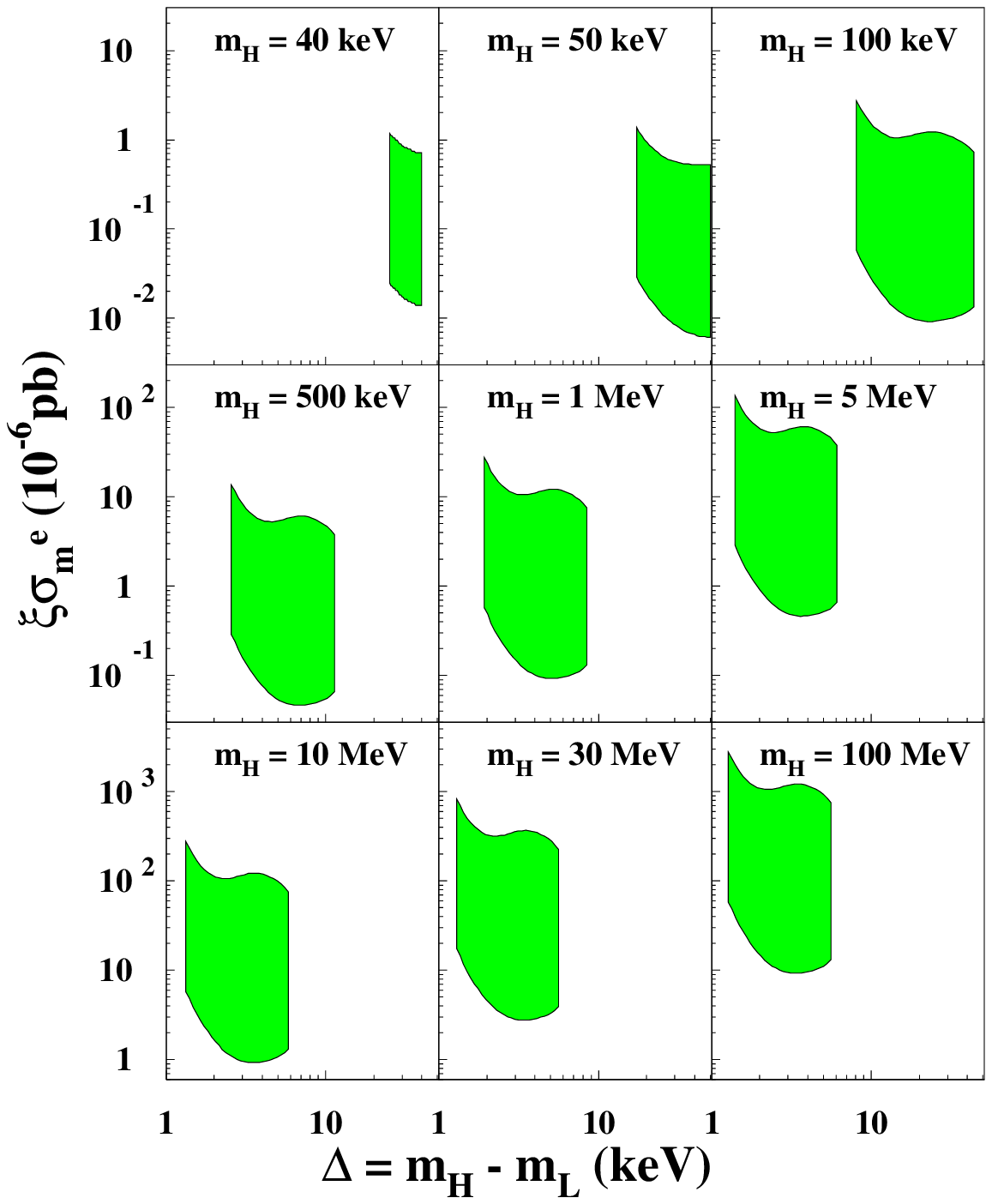}
\raisebox{2cm}{
\includegraphics[width=150pt] {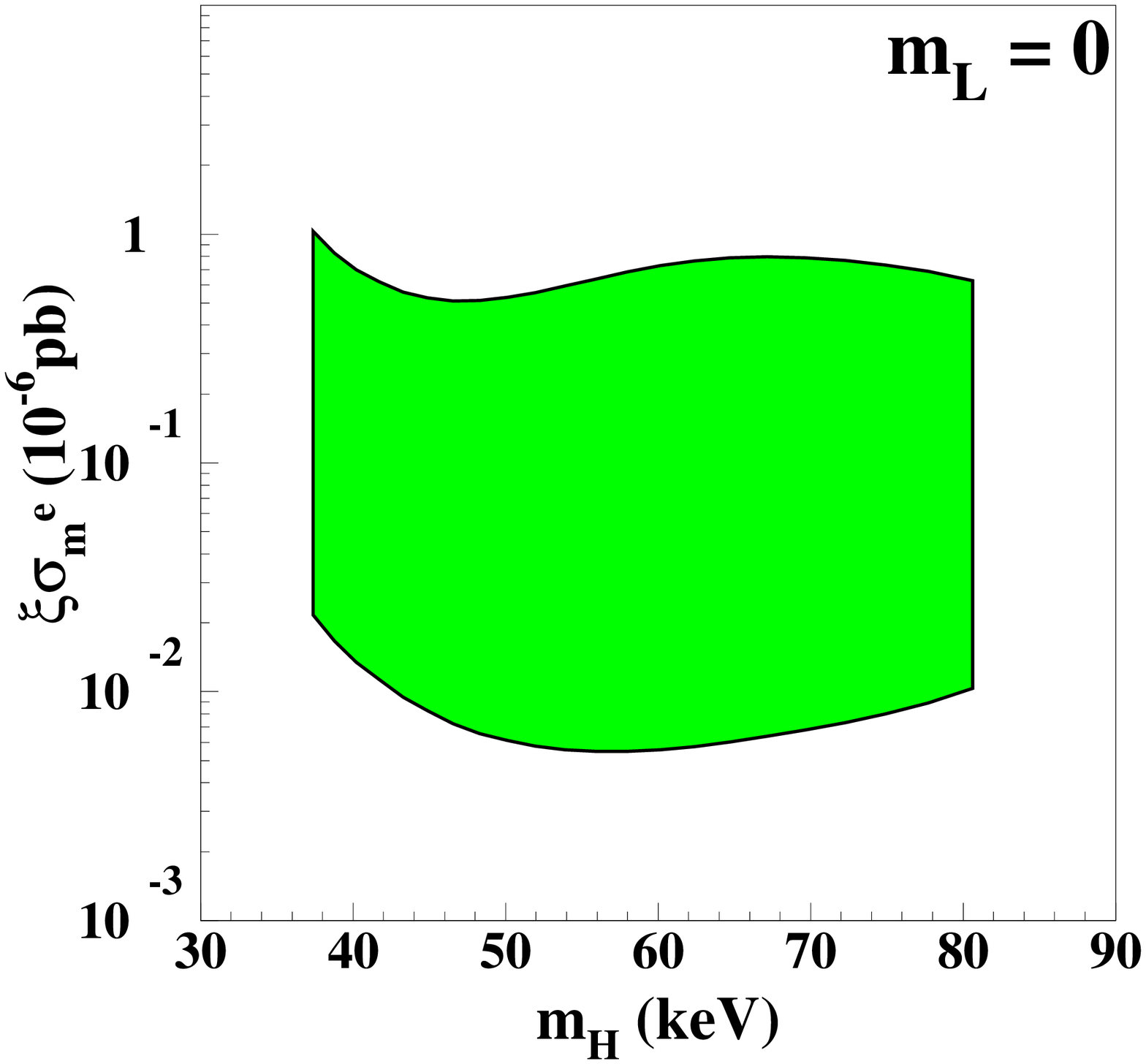}}
\vspace{-0.4cm}
\caption{Case of electron interacting LDM.
{\em Left:} examples of some slices of the 4$\sigma$ allowed 3-dimensional volume 
for various $m_H$ depicted in the ($\xi\sigma^e_m$ vs $\Delta$) plane.
{\em Right:} slice of the 4$\sigma$ allowed 3-dimensional volume 
for $m_H = \Delta$, that is for a massless or a very light $\nu_L$ particle,
as e.g. either an active neutrino or a nearly massless sterile one or the light axion, etc.
The same dark halo models and parameters described in ref. \cite{RNC} have been used; see text.
}
\label{fg:elett}
\end{figure}

As examples, some slices of the 3-dimensional allowed volume for various $m_H$ values 
in the ($\xi\sigma^e_m$ vs $\Delta$) plane are depicted in Fig.\ref{fg:elett}{\it --left}
in the scenario given above.

\vspace{0.4cm}
The slice of the 4$\sigma$ allowed 3-dimensional volume 
for $m_H = \Delta$ is shown in Fig.\ref{fg:elett}{\it --right}.
This slice has been taken along the dotted line of Fig.\ref{fg:elett1},
restricting $m_L \simeq 0$, that is for a massless or a very light $\nu_L$ particle,
such as e.g. either an active neutrino or a nearly massless sterile one or the light axion, etc.

\vspace{0.4cm}
Finally, it is worth to summarize that electron interacting LDM candidates 
in the few-tens-keV/sub-MeV range 
are allowed (see Figs.\ref{fg:elett1} and \ref{fg:elett}).
This can be of interest for example in the models of Warm Dark Matter particles --
such as e.g. Weakly Sterile Neutrino \cite{sterile}, Axino or Gravitino \cite{axgrav} --
or in the models where the DM is made of Moduli fields \cite{moduli}. 
Moreover, also configurations with $m_H$ in the MeV/sub-GeV range 
are allowed; similar LDM candidates (such as e.g. axino \cite{MeVax}, 
sterile neutrino \cite{MeVs},
moduli fields from string theories \cite{moduli}, 
and even MeV-scale LSP in supersymmetric theories \cite{mevlsp})
can also be of interest for the production mechanism of the 511 keV gammas from 
the galactic bulge \cite{dm511a,dm511sv}.

\subsection{Case of nucleus interacting LDM}

\begin{figure} [!ht]
\centering
\includegraphics[width=6.0cm] {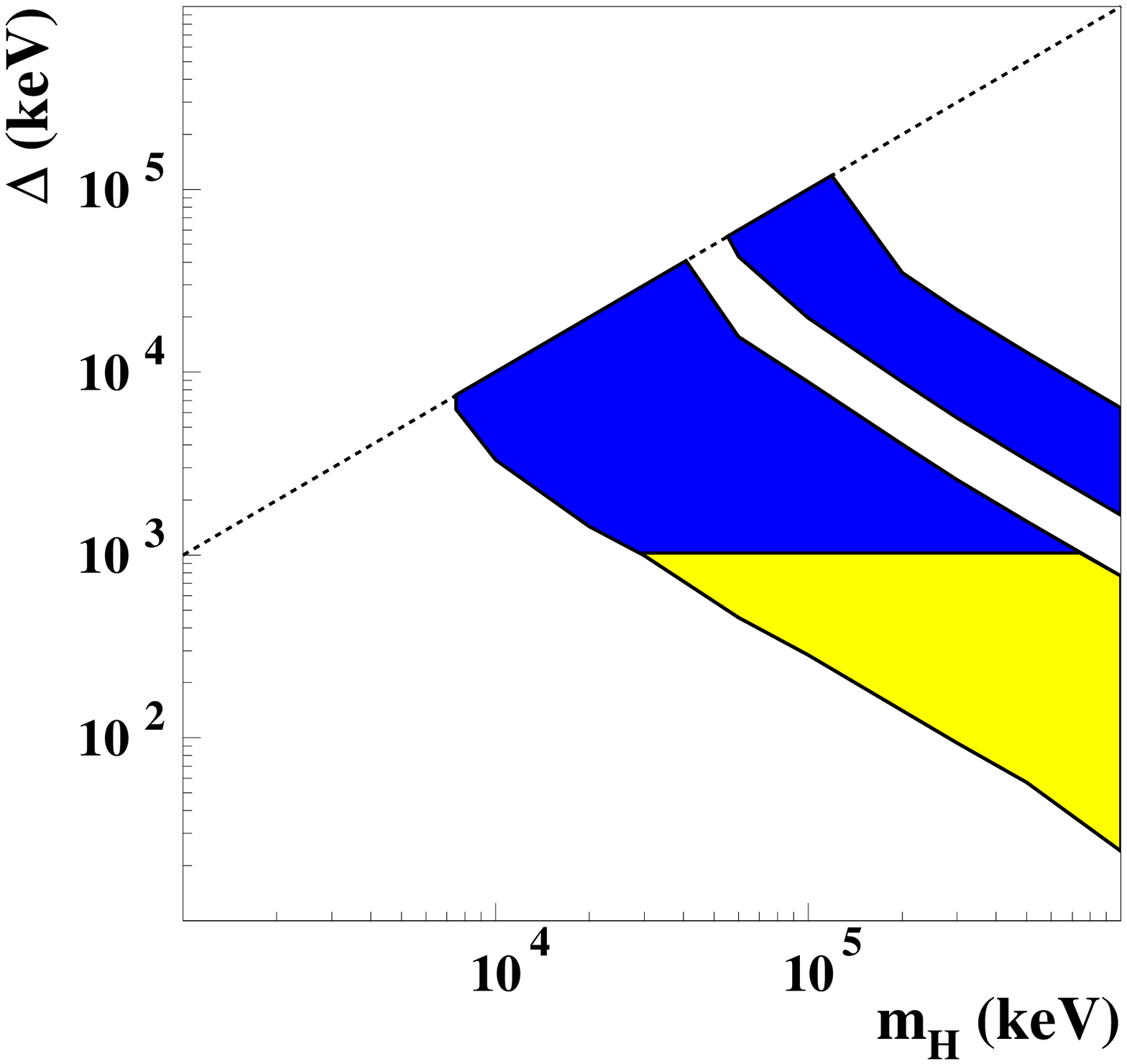}
\includegraphics[width=6.0cm] {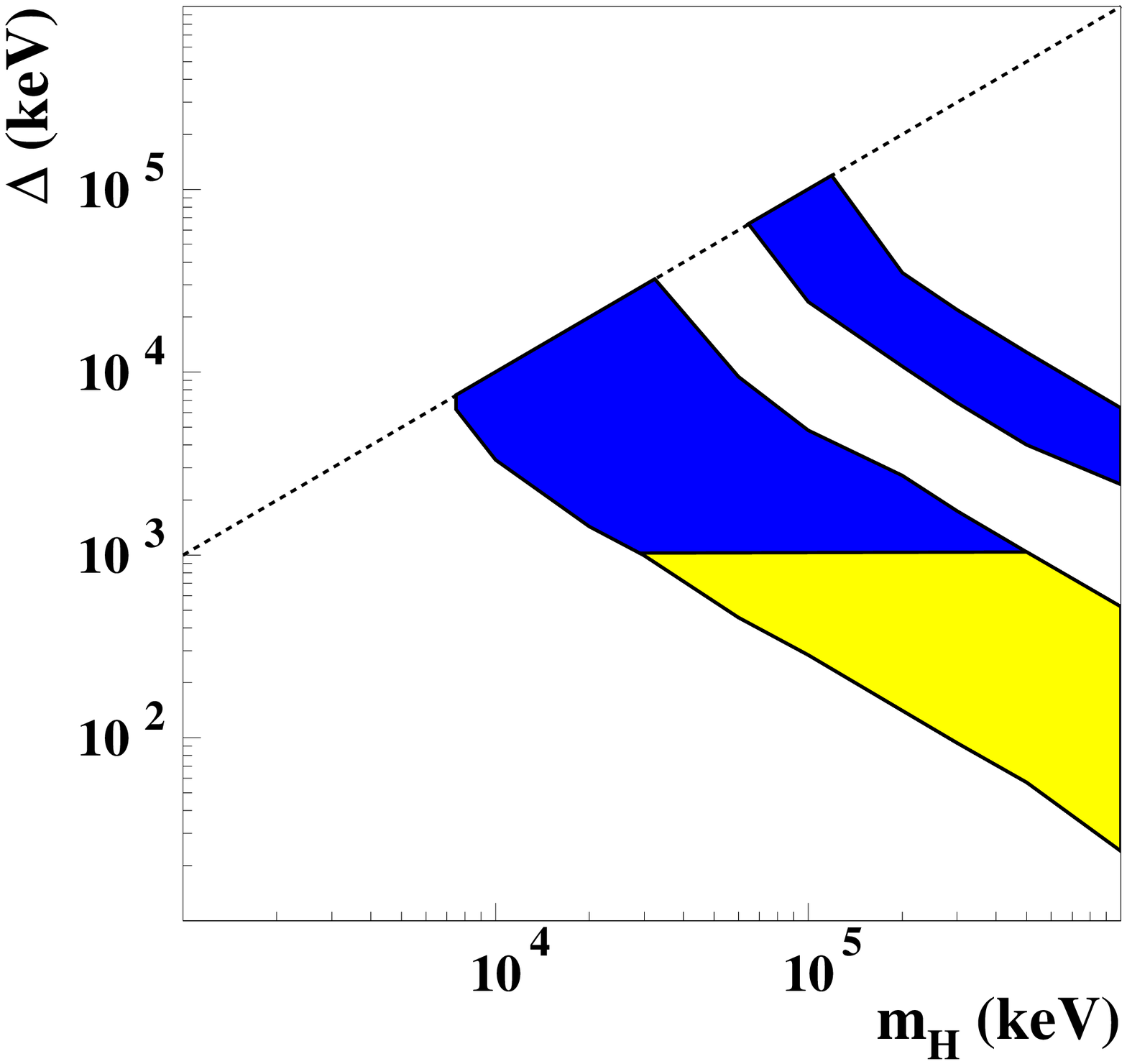}
\caption{Case of nucleus interacting LDM.
Projections of the 4$\sigma$ allowed 3-dimensional volumes on the plane ($m_H$, $\Delta$)
for the two above-mentioned illustrative cases of coherent ({\it left})
and incoherent ({\it right}) nuclear scaling laws. 
They are evaluated for the same dark halo models and parameters 
described in ref. \cite{RNC,chann}. 
The regions enclose configurations corresponding
to likelihood function values {\it distant} more than $4\sigma$ from
the null hypothesis (absence of modulation).
The dashed lines ($m_H = \Delta$) mark the case where $\nu_L$ is a massless particle.
The decays through the diagram involved in the detection channel are energetically forbidden. See text.
}
\label{fg:regn}
\end{figure}

In case of nucleus interacting LDM, 
the projections of the 4$\sigma$ allowed volumes on the plane ($m_H$, $\Delta$)
are reported in Fig.\ref{fg:regn} for the two above-mentioned illustrative cases of coherent ({\it left})
and incoherent ({\it right}) nuclear scaling laws. 
They are evaluated for the same dark halo models and parameters 
described in ref. \cite{RNC,chann}. 
The allowed $m_H$ values and the splitting $\Delta$ 
are in the intervals \mbox{8 MeV $\lsim m_H \lsim $ O(GeV)}\footnote{
We remind that for $m_H$ values greater than
O(GeV) the detection would also be possible through
the elastic scattering process \cite{RNC,ijmd,epj06,ijma2,chann}.}
and \mbox{25 keV $\lsim \Delta \lsim 120$ MeV}, respectively (see Fig.\ref{fg:regn}).
It is worth to note that in such a case the decays through the diagram involved in 
the detection channel (e.g. in nucleon anti-nucleon pairs or in meson(s), as $\nu_H \rightarrow \nu_L \pi^0$)
are obviously energetically forbidden.
Moreover, there are allowed configurations that could contribute -- in principle, if suitable
couplings exist -- to the 
positron generation in the galactic center; in fact, 
the decay $\nu_H \rightarrow \nu_L e^+ e^-$ is energetically allowed
for $\Delta > 2m_e$ (dark area in Fig.\ref{fg:regn}), 
while the annihilation processes into $e^+e^-$ pairs are energetically allowed
for almost all the allowed configurations.

It is worth to note that for nuclear interacting LDM the 3-dimensional allowed configurations 
are contained in two disconnected volumes, as seen e.g. in their projections in Fig. \ref{fg:regn}.
The one at larger $\Delta$ at $m_H$ fixed is mostly due to interaction on Iodine target, while
the other one is mostly due to interaction on Sodium target.

\begin{figure} [!ht]
\centering
\vspace{-0.4cm}
\includegraphics[width=185pt] {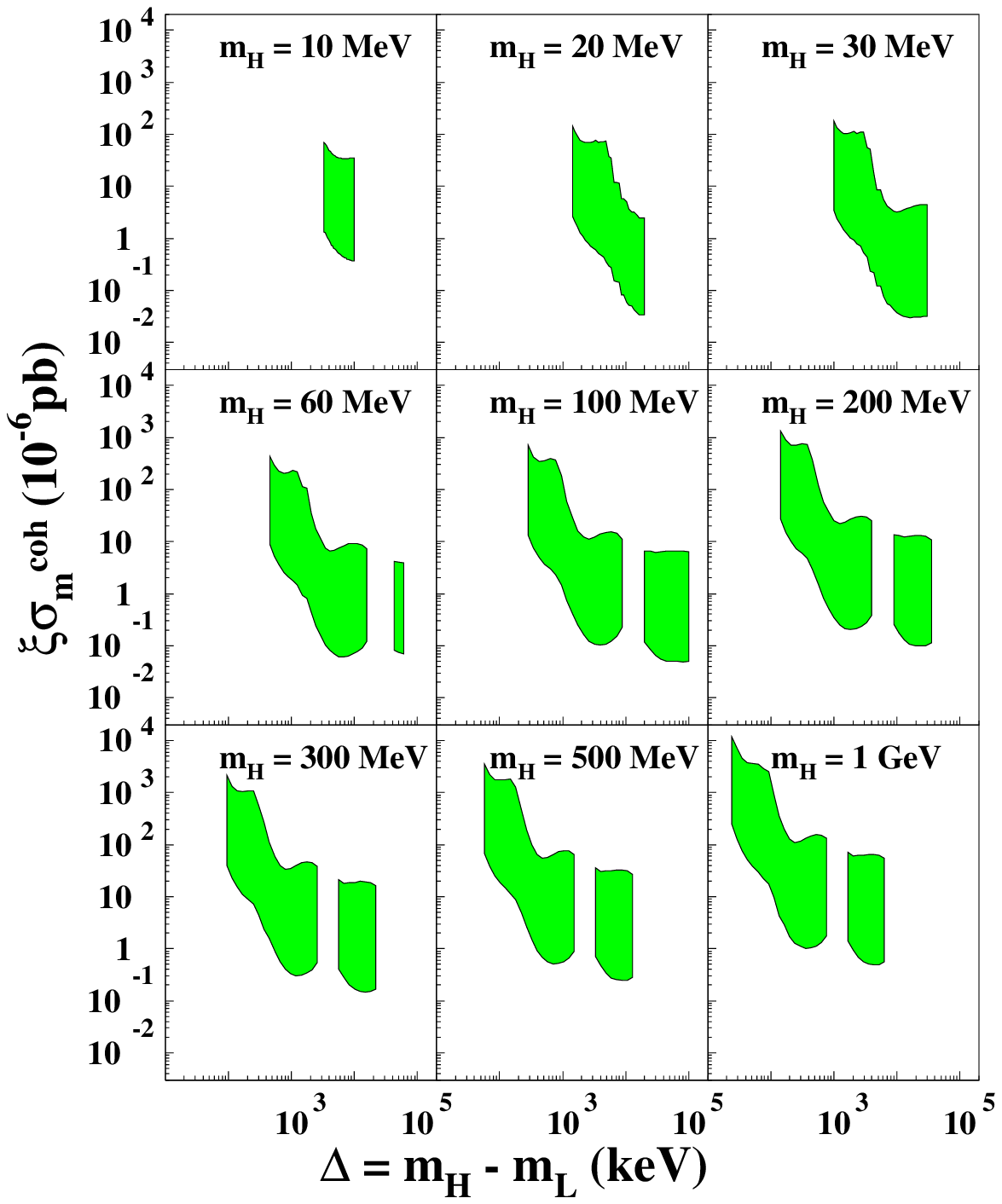}
\includegraphics[width=185pt] {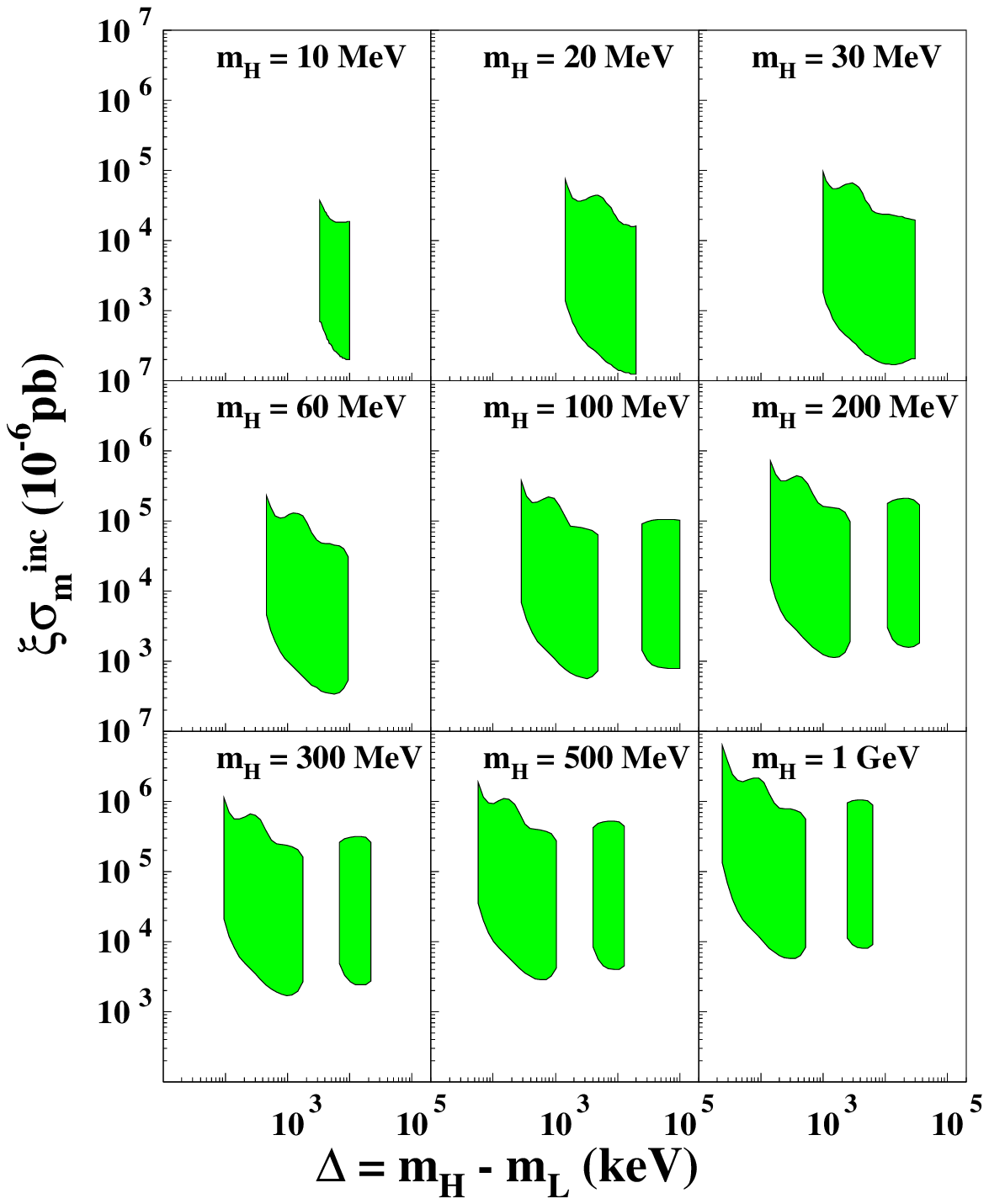}
\vspace{-0.4cm}
\caption{Case of nucleus interacting LDM.
Examples of some slices of the 3-dimensional allowed volumes for various $m_H$ values 
in the ($\xi\sigma_m$ vs $\Delta$) plane 
for the two above-mentioned illustrative cases of coherent ({\it left})
and incoherent ({\it right}) nuclear scaling laws. 
The 3-dimensional volumes enclose configurations corresponding
to likelihood function values {\it distant} more than $4\sigma$ from
the null hypothesis (absence of modulation).
The same dark halo models and parameters described in ref. \cite{RNC,chann}
have been used.}
\label{fg:panel_n}
\end{figure}

As examples, some slices of the 3-dimensional allowed volumes for various $m_H$ values 
in the ($\xi\sigma_m$ vs $\Delta$) plane are depicted in Fig.\ref{fg:panel_n}
for the two above-mentioned illustrative cases of coherent ({\it left} panel)
and incoherent ({\it right} panel) nuclear scaling laws. 

\begin{figure} [!ht]
\centering
\vspace{-0.4cm}
\includegraphics[width=140pt] {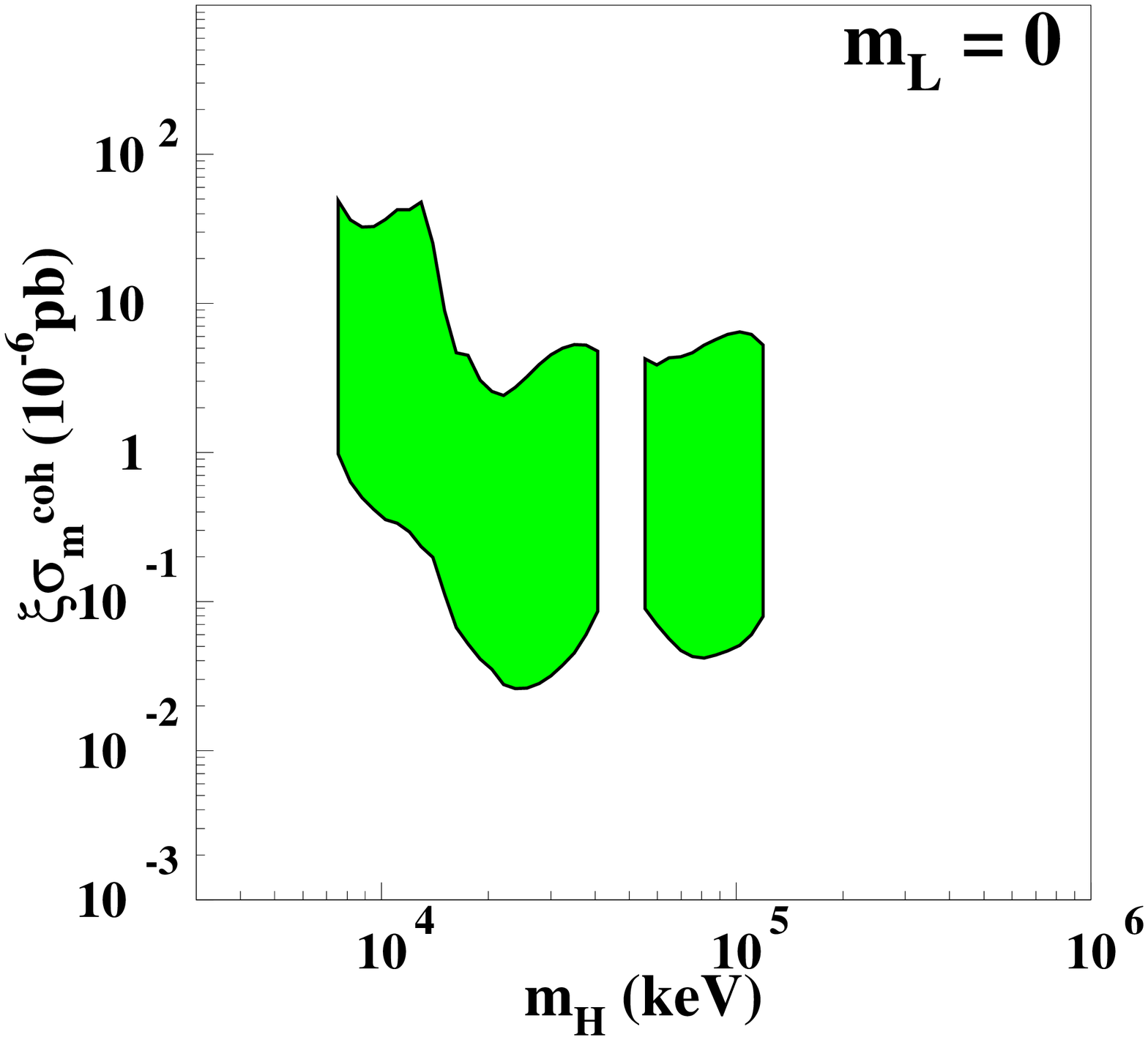}
\includegraphics[width=140pt] {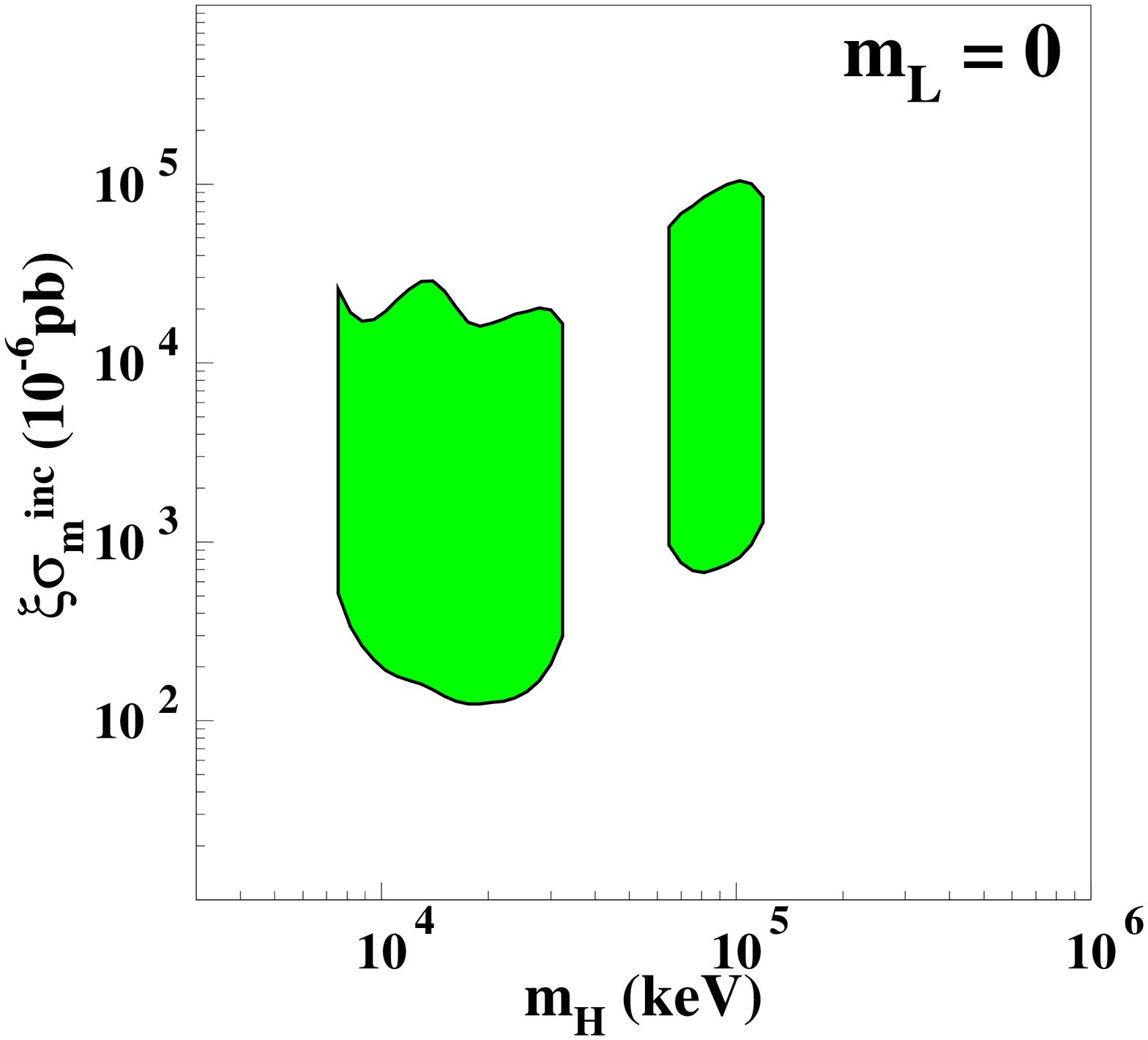}
\vspace{-0.4cm}
\caption{Case of nucleus interacting LDM.
Slices of the 4$\sigma$ allowed 3-dimensional volumes 
for $m_H = \Delta$, that is for a massless or a very light $\nu_L$ particle,
as e.g. either an active neutrino or a nearly massless sterile one or the light axion, etc.
They are evaluated for the two illustrative cases of coherent ({\it left} panel)
and incoherent ({\it right} panel) nuclear scaling laws, using 
the same dark halo models and parameters described in ref. \cite{RNC,chann}.
}
\label{fg:ster_n}
\end{figure}

The slices of the 4$\sigma$ allowed 3-dimensional volumes 
for $m_H = \Delta$ are shown in Fig.\ref{fg:ster_n}
for the two illustrative cases of coherent ({\it left} panel)
and incoherent ({\it right} panel) nuclear scaling laws. 
These slices have been taken along the dotted lines of Fig.\ref{fg:regn},
restricting $m_L \simeq 0$, that is for a massless or a very light $\nu_L$ particle,
as e.g. either an active neutrino or a nearly massless sterile one or the light axion, etc.

Finally, it is worth to summarize that LDM candidates in the MeV/sub-GeV range 
are allowed (see Figs.\ref{fg:regn}, \ref{fg:panel_n} and \ref{fg:ster_n}).
Also these candidates, such as e.g. axino \cite{MeVax}, sterile neutrino \cite{MeVs},
moduli fields from string theories \cite{moduli}, 
and even MeV-scale LSP in supersymmetric theories \cite{mevlsp},
can be of interest for the positron production in the galactic center
\cite{dm511a,dm511sv}.

\section{Conclusions}

In this paper, the possibility of direct detection of a Light Dark Matter candidate
particle has been investigated.
The inelastic scattering processes on the electron or on the nucleus targets have been considered;
these are the only possible processes useful for the direct detection of such LDM candidates.

Some theoretical arguments have been developed and related phenomenological aspects have been discussed.
In particular, the impact of the LDM candidate has also been analyzed in a phenomenological
framework on the basis of DAMA/NaI annual modulation data.
Allowed volumes and regions for the characteristic phenomenological 
parameters have been derived in the considered model.
The allowed phenomenological parameters (masses and cross sections) of LDM can be of interest
for various LDM candidates proposed in theories beyond the Standard Model
and for the production mechanism of the 511 keV gammas from the galactic bulge. 

In conclusion this paper has shown that -- in addition to other candidates, as 
WIMP/WIMP-like particles and axion-like bosons, already discussed by DAMA collaboration elsewhere
\cite{ijma,RNC,ijmd,epj06,ijma2,chann,wele} -- there is also possibility for a LDM candidate
interacting either with the electrons or with the nuclei to account for the 6.3$\sigma$
C.L. model independent evidence for the presence of a particle DM component in the galactic halo.


\begin{thebibliography}{99}
\itemsep -2pt

\bibitem{sterile} see e.g.: A. Kusenko, AIP Conf. Proc. 917 (2007) 58; \\
A. Palazzo et al., Phys. Rev. D 76 (2007) 103511; \\
M. Shaposhnikov, Nucl. Phys. B 763 (2007) 49; \\
R. Volkas, Prog. Part. Nucl. Phys. 48 (2002) 161.

\bibitem{axgrav} see e.g.: F. D. Steffen arXiv:0711.1240[hep-ph]; \\
A. Brandenburg and F. D. Steffen JCAP 0408 (2004) 008; \\
O. Seto Phys. Rev. D 75 (2007) 123506; \\
E. A. Baltz and H. Murayama JHEP 0305 (2003) 067.

\bibitem{sss} D. Hooper et al. arXiv:0704.2558[astro-ph].

\bibitem{MeVax}     D. Hooper and L. T. Wang, Phys. Rev. D 70 (2004) 063506.

\bibitem{MeVgrav}   M. Lemoine et al., Phys. Lett. B 645 (2007) 222.

\bibitem{rhn}       J. M. Fr\'ere et al., arXiv:hep-ph/0610240. 

\bibitem{MeVs}      C. Picciotto and M. Pospelov, Phys. Lett. B 605 (2005) 15.

\bibitem{moduli} 
M. Kawasaki and T. Yanagida Phys. Lett. B 624 (2005) 162;\\
T. Asaka et al., Phys. Rev. D 58 (1998) 083509;\\
T. Asaka et al., Phys. Rev. D 58 (1998) 023507.

\bibitem{elko} 
D. V. Ahluwalia-Khalilova and D. Grumiller Phys. Rev. D 72 (2005) 067701; \\
D. V. Ahluwalia-Khalilova and D. Grumiller JCAP 07 (2005) 012.

\bibitem{dm511a} 
C. Boehm et al., Phys. Rev. Lett. 92 (2004) 101301;\\
P. Fayet, Phys. Rev. D 75 (2007) 115017.

\bibitem{dm511sv} 
C. Boehm and Y. Ascasibar, Phys. Rev. D 70 (2004) 115013;\\
Y. Ascasibar et al., Mon. Not. R. Astron. Soc. 368 (2006) 1695;\\
C. Jacoby and S. Nussinov JHEP 05 (2007) 17;\\
P. Fayet Phys. Rev. D 70 (2004) 023514. 

\bibitem{dm511d}
D. P. Finkbeiner and N. Wiener, Phys. Rev. D 76 (2007) 083519;\\
J. A. R. Cembranos and L. E. Strigari, arXiv:0801.0630[astro-ph].

\bibitem{mevlsp} D. Hooper and K.M. Zurek arXiv:0801.3686 [hep-ph].

\bibitem{ijma}     R. Bernabei et al., Int. J. Mod. Phys. A 21 (2006) 1445.
\bibitem{RNC}      R. Bernabei el al., La Rivista del Nuovo Cimento 26 n.1 (2003) 1-73.
\bibitem{ijmd}     R. Bernabei el al., Int. J. Mod. Phys. D 13 (2004) 2127.
\bibitem{epj06}    R. Bernabei et al., Eur. Phys. J. C. 47 (2006) 263.
\bibitem{ijma2}    R. Bernabei el al., Int. J. Mod. Phys. A 22 (2007) 3155-3168.
\bibitem{chann}    R. Bernabei et al., Eur. Phys. J. C 53 (2008) 205.
\bibitem{wele}     R. Bernabei el al., Phys. Rev. D 77 (2008) 023506.

\bibitem{Bo03}     A. Bottino et al., Phys. Rev. D 67 (2003) 063519; \\
                   A. Bottino et al., Phys. Rev. D 68 (2003) 043506.
\bibitem{Bo04}     A. Bottino et al., Phys. Rev. D 69 (2004) 037302.
\bibitem{Botdm}    A. Bottino et al., Phys. Lett. B 402 (1997) 113; \\
                                      Phys. Lett. B 423 (1998) 109; \\
                                      Phys. Rev. D 59 (1999) 095004; \\
                                      Phys. Rev. D 59 (1999) 095003; \\
                                      Astrop. Phys. 10 (1999) 203; \\
                                      Astrop. Phys. 13 (2000) 215; \\
                                      Phys. Rev. D 62 (2000) 056006; \\
                                      Phys. Rev. D 63 (2001) 125003; \\
                                      Nucl. Phys. B 608 (2001) 461.
\bibitem{khlopov}  K. Belotsky, D. Fargion, M. Khlopov and R.V. Konoplich, hep-ph/0411093.
\bibitem{Wei01}    D. Smith and N. Weiner, Phys. Rev. D 64 (2001) 043502; \\
                   D. Tucker-Smith and N. Weiner, Phys. Rev. D 72 (2005) 063509.
\bibitem{foot}     R. Foot Phys. Rev. D 69 (2004) 036001.
\bibitem{Saib}     S. Mitra, Phys. Rev. D 71 (2005) 121302(R).
\bibitem{droby}    E.M. Drobyshevski et al., Astron. \& Astroph. Trans. 26:4 (2007) 289; \\
                   E.M. Drobyshevski, arXiv:0706.3095.
\bibitem{sneu}     C. Arina and N. Fornengo, arXiv:0709.4477.
\bibitem{zoom}     A. Bottino et al., arXiv:0710.0553.

\bibitem{sigmav}   G. Jungman et al., Phys. Rep. 267 (1996) 195; \\
                   M.A. Amin and T. Wizansky arXiv:0710.5517.

\bibitem{Ext}      P. Belli et al., Phys. Rev. D61 (2000) 023512.
\bibitem{loc}      P.J.T. Leonard and S. Tremaine, {\em Astrophys. J.} 353 (1990) 486; \\
                   C.S. Kochanek, {\em Astrophys. J.} 457 (1996) 228; \\
                   K.M. Cudworth, {\em Astron. J.} 99 (1990) 590.

\end{thebibliography}
\end{document}